\documentstyle[prl,aps,floats,psfig]{revtex}
\begin{document}
\twocolumn
 \wideabs{
\title{Width of the zero-field superconducting resistive transition
in the vicinity  of the localization threshold}

\author{V.F.Gantmakher and M.V.Golubkov}

\address{Institute of Solid State Physics RAS, Chernogolovka, 142432
Russia}

\maketitle

\begin{abstract}

Resistive superconducting zero-field transition in amorphous In-O
films in states from the vicinity of the insulator-superconductor
transition is analyzed in terms of two characteristic
temperatures: the upper one, $T_{c0}$, where the finite amplitude
of the order parameter is established and the lower one, $T_c$,
where the phase ordering takes place. It follows from the
magnetoresistance measurements that the resistance in between,
$T_c<T<T_{c0}$, cannot be ascribed to dissipation by thermally
dissociated vortex pairs. So, it is not
Kosterlitz-Thouless-Berezinskii transition that happens at $T_c$.

\end{abstract}
}

The resistive superconducting ($s$-)transition in bulk
conventional superconductors is very narrow. The reduced width
 $t\equiv\mid T-T_{c0}\mid/T_{c0}$ of the region with
strong fluctuations around transition temperature $T_{c0}$ is
$t\propto(T_{c0}/\epsilon_F)^4$ in clean limit and
$t\propto(T_{c0}/\epsilon_F)(k_Fl)^{-3}$ in dirty limit, with the
product of Fermi wavevector by mean free path $k_Fl>1$. It is
different in 2D where free magnetic vortices serve as thermal
fluctuations. Broad $s$-transitions in films were explained by
existence of the temperature range where current dissipation is
due to these fluctuations \cite{Diss}. Transition starts at
temperature $T_{c0}$, when Cooper pairs appear in the electronic
spectrum.  Below  $T_{c0}$ the resistance remains finite because
of free vortices.  They appear with probability $\mu(T)$  while
inbinding of vortex-antivortex pairs (magnetic loops).  When
external magnetic field is zero, the system of thermal
fluctuations contains equal numbers $N_+(T)=N_-(T)=N(T)$ of free
vortices of opposite signs.  Each vortex lives independently until
it annihilates after collision with a vortex of opposite sign.
Annihilation probability $\tau^{-1}(T)$ together with probability
$\mu(T)$ determine through dynamic equilibrium the concentration
$N(T)$:
\begin{equation}  \label{1}
N^2(T)=a(\mu\tau)^{-1},\qquad a={\rm const}.
\end{equation}
Assuming that there is no pinning, the resistance $R$ is
proportional to the total concentration $2N$ of the vortices:
\begin{equation} \label{R}
R=2\pi\xi_c^2(2N)R_n,
\end{equation}
with $\xi_c$ being the effective radius of the vortex core and
$R_n$ being the resistance in the normal state. The finite
resistance exists until Kosterlitz~-- Thouless~-- Berezinskii (KTB)
transition \cite{KTB} inside the vortex system takes place at some
temperature $T_{c}$. Below $T_{c}$, practically all vortices are
bound into loops and $N=0$. As loops do not dissipate energy, the
resistance vanishes at $T_c$.

This scheme with two characteristic temperatures was very
carefully checked several times with different materials. In
particular, Hebard et al. \cite{Heb} in experiments with
amorphous InO$_x$ films with $T_{c0}\approx2.5$\,K and
$T_c\approx1.8$\,K have confirmed the transport characteristics
predicted by the theory.

In the frame of the BCS theory applied to 2D, both regions
controlled by fluctuations, below and above $T_{c0}$, are narrow
differing only by a numerical factor \cite{Lark1}
\begin{equation} \label{Gi}
(T_{c0}-T_{c})/T_{c0}\approx3(T-T_{c0})/T_{c0}\approx3 {\rm Gi}\ll1,
\end{equation}
with Ginzburg parameter Gi being usually small, ${\rm Gi}\ll1$.
When the disorder is strong so that the mean free path $l$ reaches
its minimum value of $k_F^{-1}$, the Gi increases and becomes of
the order of unity, and the 2D \ KTB-transition temperature $T_c$
is suppressed compared to $T_{c0}$ so that the region
$T_c<T<T_{c0}$ widens \cite{Lark1}. Vicinity of the
superconductor-insulator ($s{-}i-$)transition is just such region.
It is tempting to describe the main part of the broad resistive
$s$-transitions in terms of vortex-induced dissipation in this
case too. However, experiments with granular Pb films \cite{Pb}
demonstrated that the scheme was not universal: the width of the
zero-magnetic-field $s$-transition for the states from the
vicinity of the $s{-}i-$transition was controlled not by thermally
activated free vortices.

The width of the fluctuation region $t$ above $T_{c0}$ increases
along with $k_Fl$ approaching unity: strong disorder makes the
fluctuation region wide. This happens not only in 2D \cite{Ind2D}
but in 3D as well \cite{Kap,Bul} so that specific properties of
``short'' vortices in 2D are here not of decisive importance.
Recent approaches for 3D \cite{EK} also distinguish between
fluctuations of the amplitude of the order parameter and those
which destroy long-range phase coherence. In such interpretation,
mean amplitude becomes finite in the vicinity of $T_{c0}$ and the
long-range phase coherence establishes at $T_c<T_{c0}$. This
problem is not yet well understood. Recently Valles et al.
\cite{Valles} concluded from tunneling measurements on ultrathin
$s$-films near $s{-}i-$transition that fluctuations in the
amplitude of the superconducting order parameter dominated below
$T_{c0}$ .

In the Ref.\,\cite{Pb}, the vortex-determined-dissipation scheme
was questioned for granular material. Here we study the same
problem for amorphous films where the disorder is supposed to be
on the atomic scale. Our amorphous InO$_x$ films were 200\,\AA\
thick. They were certainly 2D from the viewpoint of vortex
electrodynamics since the magnetic penetration depth
$\lambda\gtrsim 1000$\,\AA. The 2D character of our films becomes
not so obvious when the thickness $d$ is to be compared with some
other lengths: various estimates give for the superconducting
coherence length $\xi_0$ the value in the range 100--500\,\AA\ and
the magnetic length $l_B=(\hbar c/eB)^{-1/2}$ is 200\,\AA\ at
$B=1$\,T. The films are 3D in normal state, as the thickness $d$
is certainly larger then $l\approx k_F^{-1}$.

\begin{figure}
\vbox{\psfig{figure=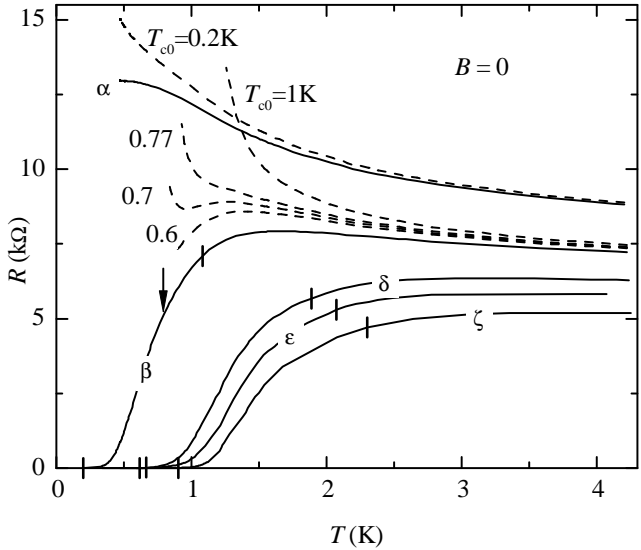,width=\columnwidth,clip=}
\caption{Zero-field resistive s-transitions for several states of
an amorphous In-O film. Bars frame the transition region
($0.9R_{\max}-10^{-3}R_{\max}$) on the curves $R(T)$. Dash
curves~-- one for state $\alpha$ and several, with different
values of parameter $T_{c0}$, for state $\beta$~-- show virtual
normal state resistance $R_n^*(T)$ obtained after subtraction of
the paraconductivity term calculated in accordance with
Eq.(\protect\ref{ALa}). The selected value of $T_{c0}$ for state
$\beta$ is shown by the arrow.  } }
\end{figure}

The properties of the film are determined by the oxygen
concentration $x$ \cite{Ova}. The starting value of $x$ can be
changed in some extend by thermal treatment. This affects the
carrier concentration and the position of the state on the
$s{-}i-$phase diagram [11\,--\,13]. We remain in the region where
the carrier density $n$ judging from Hall effect measurements is
in the range $(2-4)\cdot 10^{21}$cm$^{-3}$ and the parameter
$k_Fl$ is in the range $0.2-0.3$ \cite{Ova}. Hence, in terms
introduced by Emery and Kivelson \cite{EK}, we deal with a ``bad''
(non-Drude) metal, where the transport phenomena  are not
described by Boltzmann theory.

In the experiments, resistive $s$-transition $R(T,B)$ is measured.
Below, data for several states of one film are demonstrated.
Results for other films are similar. The aspect ratio of the film
is close to one: its resistance $R$ serves within 10\% accuracy as
resistance per square. The measurements were done for 6 states of
the film labeled as $\alpha, \beta,\gamma,
\delta,\varepsilon,\zeta$ with $s$-properties gradually increasing
along this row. Fig.1 contains functions $R(T)$ in zero magnetic
field for five of these states. In state $\alpha$, the
s-transition, if exists at all, starts somewhere below
$T\lesssim0.4$\,K. For all the others, two conditional
temperatures, $T_{c0}$ and $T_c$, are marked by bars. They may be
considered as the onset and the end of the transition.  The upper
is positioned at the level $R\approx0.9R_{\max}$ where $R_{\max}$
is the value of maximum at the curve $R(T)$. The lower is at the
level
\begin{equation}
\label{Tc} R\approx10^{-3}R_{\max},
\end{equation}
which roughly corresponds to the usual position of the KTB
transition \cite{Heb}. The problem is in factors which control the
shape of the $s$-transition in between the marks.

Under the standard approach in conventional superconductors,
$T_{c0}$ is determined from experimentally measured $R(T)$ by the
help of the expression for the paraconductivity $\sigma_{fl}$  due
to superconducting fluctuations \cite{AL}. In 2D
$$\sigma=\sigma_n+\sigma_{fl}=
\frac{e^2}{\hbar}\left[g+\frac{T_{c0}}{16(T-T_{c0})}\right], $$
\begin{equation}  \label{ALa}
\sigma=R^{-1},\quad\sigma_n=R_n^{-1}.
\end{equation}
For films far from the localization threshold, the
dimensionless sheet conductance is $g\gg1$ and the correction
to $R$ from $\sigma_{fl}$ becomes soon negligible when $T$ is
increasing above $T_{c0}$. For our films $g$  is of the order
of unity. Hence, the contribution $\sigma_{fl}$ really affects
the temperature dependence $R(T)$ above $T_{c0}$.

The term $\sigma_{fl}$ in the relation (\ref{ALa}) contains
$T_{c0}$ as the only parameter which we have to choose.
Relation (\ref{ALa}) is valid only until the correction is
small: $\sigma_{fl}\ll\sigma_n$. Hence, even with right value
of parameter $T_{c0}$ we'll get from Eq.(\ref{ALa}) function
$R_n(T)$ which falsely tends to infinity near $T_{c0}$. To
emphasize this, we'll mark these functions by star, as $R_n^*$.

Fig.1 presents functions $R_n^*$ for state $\beta$, with different
values of $T_{c0}$ as parameter in $\sigma_{fl}$. The curve
$R_n^*$ obtained with $T_{c0}=1$\,K has improbable strong
temperature dependence below 1.5\,K whereas the curves with
$T_{c0}=0.6$\,K and 0.7\,K contain maxima.  Hence, $T_{c0}$ should
be in between.

\begin{figure} [t]

\vbox{\psfig{figure=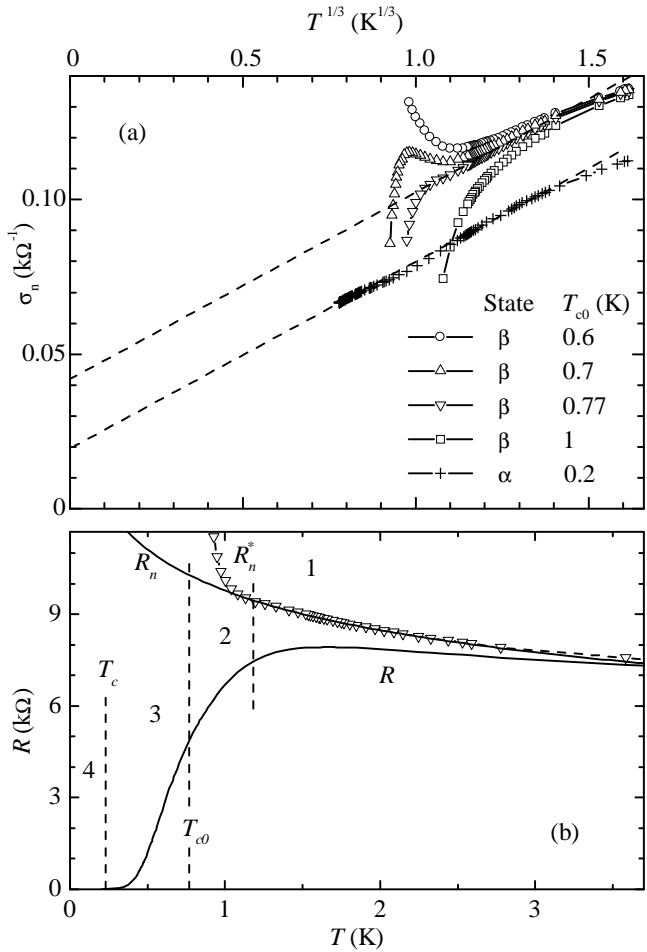,width=\columnwidth,clip=}
\caption{(a)~The dash curves from Fig.1 presented as
$\sigma_n^*=(R_n^*)^{-1}$ vs $T^{1/3}$ (form appropriated for a 3D
non-Drude metal in critical region near the metal-insulator
transition). Linear extrapolation cuts off the tail from the
region of strong fluctuations and transforms $\sigma_n^*$ into
$\sigma_n$. (b)~ Functions $R$, $R_n$, and $R_n^*$ for state
$\beta$. About four temperature regions see text.}
 \label{f1} }
\end{figure}

The specific choice of 0.77\,K as $T_{c0}$ is justified by the
plot of $\sigma_n$ vs $T^{1/3}$ (Fig.2). The representation
\begin{equation}  \label{T^1/3}
\sigma_n=u+vT^{1/3}
\end{equation}
is usually used for 3D ``bad'' metals to distinguish by
extrapolation to $T=0$ metals and insulators (see, for
instance, \cite{gMIT}). The choosed value of $T_{c0}$ gives the
lowest left-edge value of the temperature interval where data
follow Eq.(\ref{T^1/3}). For state $\beta$ with this $T_{c0}$,
the extrapolated value of $u=\sigma_n(0)$ is $0.15e^2/\hbar$.
Note that according to the definition (\ref{ALa}), $\sigma_n$
is 2D conductivity, $\sigma_n\equiv\sigma_n^{\rm (2D)}$, where as
3D conductivity is
$\sigma_n^{\rm (3D)}(0)=\sigma_n^{\rm (2D)}(0)d=(0.15d)e^2/\hbar$.

For state $\alpha$, the contribution from $s$-fluctuations is
clearly seen in Fig.1 as tendency to decline at low temperatures.
The procedures from Fig.1 applied to this state give
$T_{c0}=0.2$\,K:  this is the lowest value of parameter $T_{c0}$
which brings the curve $R_n^*(T)$ without maximum.  According to
Fig.2a, the extrapolated value of $\sigma_n(0)$ for state $\alpha$
is twice as small as for state $\beta$. One more such step should
bring the system to the localization threshold. We know from \cite
{Ova} that this would result in zero-field $s{-}i-$transition.

Returning to state $\beta$, the kink on the curve
$\sigma_n(T^{1/3})$ in Fig.2a reveals the point where fluctuations
become so strong that Eq.(\ref{ALa}) fails. As it is shown in
Fig.2b, the temperature axes breaks out into four regions. In the
right one, the paraconductivity exists. In region 2, strong
superconducting fluctuations prevail. At the opposite end, region
4 is superconducting. Our next task is to study region 3 and to
check whether it is the vortex dissipation that controls the
resistance in this region, i.e. in the lower part of the
transition.

Let us turn to isotherms $R(B)$ on Fig.3. All the states studied
are situated on the $s$-side of the phase diagram \cite{g1} of the
$s{-}i-$transition. Certain critical field values of $B_c$ induce
$s{-}i-$transition in these states and bring the sample in to the
intermediate position between the superconductor and the insulator
\cite{Fish}. The resistance at this field, $R(T,B_c)=R_c$, should
not depend on temperature at all \cite{Fish} or may have only weak
temperature dependence \cite{g2}.  Hence the isotherms $R(B,T={\rm
const})$ cross in the vicinity of $B_c$. At low fields, all the
isotherms from the vicinity of $T_c$ approach the origin, possibly
ending at one of the axes near the origin. Hence, the bunches in
Fig.3 have specific shape of lenses.

Inside each lens, one can more or less confidently select some
mean isotherm which separates those with different signs of the
second derivative $\partial^2R/\partial B^2$ inside the interval
$0\div B_c$. Corresponding temperatures of these separating
isotherms are written near the bunches. For the left bunches
$\gamma$ and $\delta$, the separating isotherms turn to be
straight lines in the aforementioned interval with the slope
$\partial R/\partial B\approx R_c/B_c$. For the states
$\varepsilon$ and $\zeta$ situated deeper in the $s$-region, the
lenses are slightly deformed and the separating isotherms remain
straight only below 2--3\,T.

The temperatures of the separating isotherms practically coincide
with the values of $T_c$ determined by criterion (\ref{Tc}). To
some extend, this justifies the choice of the coefficient in the
criterion (\ref{Tc}). On the other hand, we get a more convenient
tool to determine the temperature where the $s$-transition becomes
complete: by finding the isotherm $R_{T_c}(B)$ which is linear
function going through the origin.

\begin{figure} [h]
\vbox{\psfig{figure=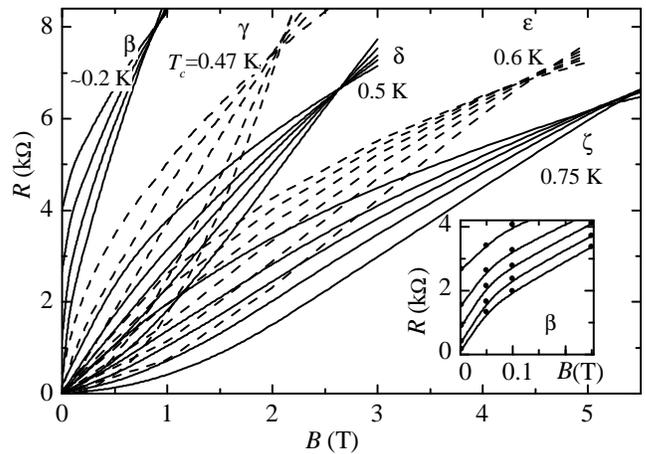,width=\columnwidth,clip=}
\caption{Magnetoresistance isotherms for a sequence of temperature
values for several states of the In-O film.  The temperatures in
each bunch are downward from 0.85~K with step 0.1~K. About the
labeled values of $T_c$ see text. Inset:  more dense set of
isotherms for state $\beta$, downward from 0.58~K with step 0.05~K
}}
\end{figure}

The isotherms with $T>T_c$ cross the ordinate at finite
$R(0)\neq0$. It is  clear from the specific shape of the low-field
part of the lenses that they have nonzero slope at $B=0$ (detailed
demonstration can be found in the inset in Fig.3):

\begin{equation}  \label{slope}
(\partial R/\partial B)_0>0.
\end{equation}
This linear increase of $R$ exists only with the field
perpendicular to the film. The field directed along the film
which does not bring the vortices from outside into the film
results in zero field derivative $(\partial R/\partial
B)_0=0$, i.e. does not affect dissipation in the linear
approximation. This can be seen from the previously published
data on In-O films (Fig.2 in \cite{Annal}) which compare
isotherms for two almost similar states of the film but with
different directions of the applied magnetic field.

In the model of thermally excited free vortices, the
normal applied field increases the density of the vortices of
the corresponding sign: $\Delta N_+\propto B$.  But this leads,
due to the recombination processes, to lessening of the density
of vortices of opposite sign. When
\begin{equation}  \label{Delta-nu}
\Delta R\equiv(R(B)-R(0))\ll R(0), ~~\mbox{i.e.}~~
\Delta N_+\ll N\:,
\end{equation}
it follows from the dynamic equilibrium equation (\ref{1}) that
$(N+\Delta N_+)(N+\Delta N_-)=N^2$, i.e. that in the linear
approximation the density changes compensate each other: $\Delta
N_-=-\Delta N_+$. This qualitative inference illustrates result
calculated by Minnhagen \cite{Minn} far ago: the free vortex
density did not change until relation (\ref{Delta-nu}) was valid.

Hence, in the frame of free vortex model, one should expect the
resistance change under the condition (\ref{Delta-nu}) to be
$\Delta R=O(B^2)$. Experimental observation that $\Delta R\propto
B$ when $\Delta R\ll R(0)$ means that the zero-field resistance
$R$ at this temperature is determined not by vortices from thermally
dissociated  pairs.

Summarizing, we described the upper part of the resistive
$s$-transition of a ``bad'' (non-Drude) metal, amorphous In-O
film, by the usual expression \cite{AL} for 2D paraconductivity
but failed to describe  the lower part in terms of a $s$-material
with thermally excited vortices. Being in lines with the idea of
two characteristic temperatures, our analysis does not confirm
existence of KTB transition in the vicinity of the lower one.
According to Emery and Kivelson \cite{EK}, separation of onset
temperature $T_{c0}$ where the amplitude of the order parameter is
established and of the phase ordering temperature $T_c<T_{c0}$
happens also in 3D non-Drude metals.  Here the
phase-order-breaking thermal fluctuations are certainly not
vortices. Our experiment seems to be closer to this model.
\vspace{3cm}

We thank V.\,T. Dolgopolov, A.\,A. Shashkin, and G.\,E. Tsydynzhapov
for
discussions and cooperation in experiments. V.\,F.\,G. thanks Weizmann
Institute of Science, Israel, for hospitality and A. Finkelstein,
Y. Levinson, M. Schechter and Y. Oreg for fruitful discussions during
his visit when the final version of this paper was written.

\end{document}